\documentclass[shortnotes,twocolumn]{jpsj2} 

\newcommand{\msr}{$\mu$SR}

\title{Nonlocal effect on the magnetic penetration depth in multigapped superconductors}

\author{%
Ryosuke \textsc{Kadono}$^{1, 2}$}%
\inst{$^{1}$Muon Science Laboratory, Institute of Material Structure
Science, High Energy Accelerator Research Organization (KEK), 1-1 Oho,
Tsukuba, Ibaraki 305-0801\\
$^{2}$Department of Materials Structure Science, The Graduate University
for Advanced Studies (SOKENDAI), Tsukuba, Ibaraki 305-0801\\
}

\abst{%
}

\kword{superconductivity, nonlocal effect, multigap, MgB$_2$}

\begin{document}
\maketitle

Nonlocal effect is of fundamental importance to the understanding of superconductivity, 
as it plays a key role in various physical phenomena observed in superconductors.
In general, the term ``nonlocal" bears a meaning that the electromagnetic response of
superconductors is not described by local equations due to the fact that the Cooper
pairs carrying the supercurrent have a finite distance of correlation (the ``coherence length", $\xi_0$). 
Meanwhile, the same term often refers to the situation that the physical quantities 
controlling superconductivity depends on electronic momenta.  
In this short note, we discuss the nonlocal 
effect in the former sense, where it is pointed out that the effective magnetic penetration depth
($\lambda$) in the multigapped superconductors may exhibit  
decrease with decreasing external field ($B_0$) in a manner 
$\lambda\propto\sqrt{B_0/B^*_{c2}}$ (with $B^*_{c2}$ being the 
upper critical field for the smallest energy gap)
due to the nonlocal effect.

In type I superconductors, it is well established that the nonlocal effect 
leads to the enhancement of the penetration depth,
\begin{equation}
\lambda\simeq(\lambda_{\rm L}^2\xi_0)^{1/3},\label{mt}
\end{equation}
where $\lambda_{\rm L}$ corresponds to the local London penetration depth.\cite{Tinkham:96}
The above estimation is obtained from the elementary argument for a bulk superconductor,
considering a mean vector potential ($\overline{A}=\lambda B_0$) over a surface 
layer of thickness $\lambda$ (which is zero elsewhere).  Here, let us follow the 
similar line to clarify the situation in type II superconductors.  
We introduce a vector potential
for an isolated flux line.
\begin{equation}
A(r)=\frac{\Phi_0}{2\pi r},\label{vecpot}
\end{equation}
where $\Phi_0$ is the flux quantum and $r$ is the distance from the center of the flux.
The above equation may be approximated into a mean potential,
\begin{equation}
\overline{A}\simeq\frac{\Phi_0}{2\pi \lambda}\simeq\frac{SB_0}{2\pi \lambda}
\end{equation}
within a cylinder ($0\ll r\le\lambda$) and zero elsewhere (with $S$ being the
unit surface of flux line lattice).
The average current density $\overline{J}$ around the flux can be estimated by using the nonlocal
form of the supercurrent and $\overline{A}$,
\begin{equation}
\overline{J}\simeq-\frac{c}{4\pi\lambda_{\rm L}^2}\frac{\lambda^2}{\xi_0^2}\overline{A}
=-\frac{c\lambda}{4\pi\lambda_{\rm L}^2}\frac{SB_0}{2\pi\xi_0^2},\label{curr}
\end{equation}
where the factor $\lambda^2/\xi_0^2$ comes from the ratio of volume for  
the region of non-zero $\overline{A}$ ($\propto\pi\lambda^2$) 
to the effective integration volume  ($\propto\pi\xi_0^2$).
Combining eq.~(\ref{curr}) with a Maxwell equation around the flux, we have
\begin{equation}
\frac{B_0}{\lambda}\simeq|\langle {\rm curl} {\bf B}\rangle | = \frac{4\pi \overline{J}}{c}
\simeq\frac{\lambda}{\lambda_{\rm L}^2}\cdot\frac{SB_0}{2\pi\xi_0^2},
\end{equation}
from which we obtain a solution
\begin{equation}
\lambda\simeq\lambda_{\rm L}\sqrt{\frac{2\pi\xi_0^2}{S}}.\label{lmd}
\end{equation}
Using an expression for the upper critical field
\begin{equation}
B_{c2}=\frac{\Phi_0}{2\pi\xi_0^2}=\frac{SB_0}{2\pi\xi_0^2},
\end{equation}
eq.~(\ref{lmd}) is rewritten as
\begin{equation}
\lambda\simeq\lambda_{\rm L}\sqrt{\frac{B_0}{B_{c2}}}.
\end{equation}
Thus, when the nonlocal effect is significant (i.e., $\lambda_{\rm L}\leq\xi_0$), 
the penetration depth around flux line is renormalized by a factor $\sqrt{B_0/B_{c2}}$.
Interestingly, the effect discussed here serves to {\sl reduce} the penetration depth,
which is in marked contrast to the case of eq.~(\ref{mt}).

It would be needless to mention that the situation considered in the above estimation 
cannot be realized in ordinary type II superconductors with a single gap, because the 
condition of $\lambda_{\rm L}\leq\xi_0$ is never satisfied.   However, in the case 
of multigapped supercondutors, it might happen that 
the coherence length corresponding to the smaller gap may be comparable to  
the penetration depth.  To examine the practical situation, let us discuss an
order parameter having two energy gaps, $\Delta_i$ ($i=\sigma,\pi$) with $\Delta_\pi<\Delta_\sigma$, 
 that is actually realized in MgB$_2$.\cite{Tsuda:05}
The presence of two gaps implies that there are two corresponding coherence
lengths and associated upper critical fields 
\begin{equation}
B_{c2(i)}=\frac{\Phi_0}{2\pi\xi_i^2},
\end{equation}
\begin{equation}
\xi_i\simeq\frac{\hbar v_{\rm F}}{\pi\Delta_i}\:\:\: (i=\sigma,\pi),
\end{equation}
with $v_{\rm F}$ being the Fermi velocity.
Provided that $\lambda_{\rm L}\le\xi_\pi$, the nonlocal effect 
leads to the renormalization of the penetration depth into an effective one,
\begin{equation}
\lambda\simeq\lambda_{\rm L}\sqrt{b_\pi},\label{leff}
\end{equation}
where $b_\pi\equiv B_0/B_{c2(\pi)}$.  It must be noted that 
superconductivity is maintained by the larger gap ($\Delta_\sigma$) 
for $ B_0<B_{c2(\sigma)}$,
while $\Delta_\pi$ collapses and associated nonlocal effect disappears
when $B_0>B_{c2(\pi)}$.  Meanwhile, eq.~(\ref{leff}) would not be valid for 
$\lambda<\xi_\sigma$, where $J$ is reduced to zero.
Thus, the nonlocal effect due to the 
smaller gap is predicted to occur only over the lower field range 
$\xi_\sigma^2B_{c2(\pi)}/\lambda_{\rm L}^2\le B_0<B_{c2(\pi)}$ (or 
$\xi_\sigma^2/\lambda_{\rm L}^2\le b_\pi<1$).

It is well established that muon spin rotation ($\mu$SR) measurement
in type II superconductors provides a direct information on the spatial distribution 
of magnetic field
\begin{equation}
B({\bf r})=\sum_{\bf K}b({\bf K})\exp(-i{\bf K}\cdot{\bf r}),
\end{equation}
where ${\bf K}$ are the vortex reciprocal lattice vectors. 
The Fourier component of the field profile, $b({\bf K})$, is determined
by the penetration depth and coherence length; if we adopt the London 
model, we have
\begin{equation}
b({\bf K})=\frac{B_0\exp(-\frac{1}{2}\xi_c^2K^2)}{1+\lambda^2K^2},\label{bk}
\end{equation}
where $\xi_c$ ($\propto\xi_0$) is the cutoff parameter to correct the nonlocal
electrodynamics near the vortex cores.
In a Gaussian approximation, the spin relaxation observed by \msr\  is 
described by the Gaussian decay with a relaxation rate
\begin{equation}
\sigma_\mu=\gamma_\mu\langle\sum_{\bf K}b({\bf K})^2\rangle^{1/2}\simeq \frac{G(b)}{\lambda^2},
\:\:(b\equiv B_0/B_{c2})\label{sigma}
\end{equation}
in which $\gamma_\mu$ is the muon gyromagnetic ratio, and the factor 
$G(b)\propto(1-b)[1+3.9(1-b)^2]^{1/2}$ represents the reduction of $\sigma_\mu$ mainly due to the
stronger overlap of vortices ($\propto1-b$) 
and additional decrease due to the
contribution of vortex cores at higher fields.\cite{Brandt:88}   
The nonlocal effect for the two energy gaps may be
incorporated by substituting Eq.~(\ref{lmd}) to the above equations so that 
the relaxation rate exhibits a field dependence
\begin{equation}
\sigma_\mu\simeq \begin{cases}
       G(b)/(\lambda_{\rm L}^2b_\pi), & \xi_\sigma^2/\lambda_{\rm L}^2\le b_\pi<1 \\
       G(b)/\lambda_{\rm L}^2, & 1<b_\pi<B_{c2(\sigma)}/B_{c2(\pi)}
       \end{cases}\label{sigmad}
\end{equation}
where $b=B_0/B_{c2(\sigma)}$.
It implies that $\sigma_\mu$ may exhibit strong deviation from that for 
the single gap below $B_{c2(\pi)}\simeq\Phi_0/(2\pi\xi_\pi^2)$ with a sharp 
increase with {\sl decreasing} field  ($\sigma_\mu\propto1/b_\pi$).  

The predicted behavior of $\sigma_\mu$ is surprisingly close to that found in the 
earlier reports on MgB$_2$.\cite{Niedermayer:02,Ohishi:03,Serventi:04} 
Prior to the establishment of double gap superconductivity, the strong 
enhancement of $\sigma_\mu$ at lower fields was attributed to 
uncontrolled influence of flux pinning.\cite{Niedermayer:02,Ohishi:03}
Although the flux pinning still remains as a possible cause of enhanced $\sigma_\mu$, 
we argue that the nonlocal effect discussed above may play a significant  
contribution to the observed field dependence of $\lambda$.
To visualize the situation, we reproduce the previous data\cite{Ohishi:03} and a result of
curve fitting by Eq.~(\ref{sigmad}) with $B_{c2(\pi)}$ as 
a free parameter; the data below 0.1 T were excluded from the fit as the field
dependence of  $\sigma_\mu$ was reversed probably due to the reduced flux density
near the lower critical field.  The curve is in 
excellent agreement with data with $B_{c2(\pi)}=0.53(1)$ T [$\xi_\pi=25(1)$ nm], 
reproducing the tendency of steep upturn with decreasing field below the presumed 
upper critical field for the smaller gap.  The field dependence of $\sigma_\mu$ reported by 
other groups exhibit qualitatively similar trend with a sharp upturn of 
$\sigma_\mu$ below $\sim$0.5 T, and thereby it is likely that the
relevant feature stems from an intrinsic property that might be ascribed 
to the nonlocal effect.

\begin{figure}[t]
\begin{center}
\includegraphics[width=0.85\linewidth]{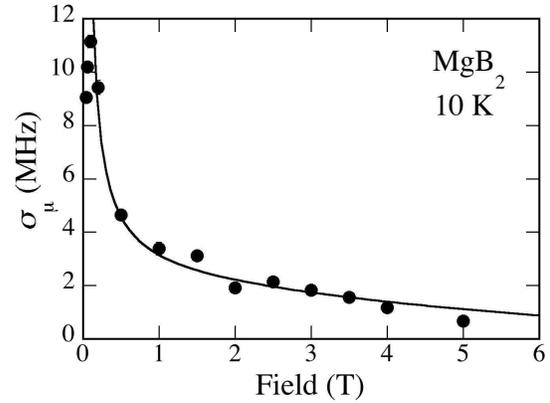}%
\caption{Muon spin relaxation rate in the superconducting state of MgB$_2$
deduced by fits using a Gaussian damping (data reproduced from Ref.\cite{Ohishi:03}).
Solid curve is a fit by the model taking account of nonlocal effect arising from 
double gap structure in the order parameter (see text for detail).
\label{sigmab}}
\end{center}
\end{figure}

The present discussion may cast some doubts to the earlier model\cite{Serventi:04} 
for explaining the behavior of $\sigma_\mu$ in MgB$_2$;  they assume that $B({\bf r})$ 
consists of a weighted sum 
of the London expression [Eq.~(\ref{bk})] with the cutoff replaced by $\xi_\pi$ and 
$\xi_\sigma$ for 
the respective components. Unfortunately, the London approximation would certainly 
fail for the smaller gap over the field range $b_\pi>0.25$.\cite{Brandt:88}

It must be noted that, while the agreement between data and the present model
in Fig.~\ref{sigmab} is encouraging, our estimation is no more than a very crude 
approximation [e.g., the use of $\overline{A}$ in Eq.(\ref{curr})] and more 
precise evaluation is clearly needed. 
Moreover, there are other sources of nonlocal effects 
that should be considered as well.  For example, recent calculation based on the quasiclassical
Eilenberger theory indicates the presence of a nonlocal effect even for the 
single gap,  where the effect may be observed as a 
gradual increase of $\lambda$ with increasing field.\cite{Laiho:07}  
A similar model for the multigap superconductors is strongly awaited for the detailed understanding
of the nonlocal effect.

\end{document}